\documentclass[lettersize,journal]{IEEEtran}

\usepackage{amsmath,amsfonts}
\usepackage{algorithmic}
\usepackage{algorithm}
\usepackage{array}
\usepackage[caption=false,font=normalsize,labelfont=sf,textfont=sf]{subfig}
\usepackage{textcomp}
\usepackage{stfloats}
\usepackage{url}
\usepackage{verbatim}
\usepackage{graphicx}
\usepackage{cite}
\usepackage{multirow}
\usepackage{booktabs}
\usepackage{float}
\usepackage{glossaries}
\usepackage{hyperref}
\usepackage{ragged2e}
\usepackage{enumitem}
\usepackage{color}
\usepackage{epstopdf}
\usepackage{hyphenat}
\def\BibTeX{{\rm B\kern-.05em{\sc i\kern-.025em b}\kern-.08em
    T\kern-.1667em\lower.7ex\hbox{E}\kern-.125emX}}
\usepackage{balance}
\newglossaryentry{DAO}{
    name={DAO},
    description={Decentralized Autonomous Organization}
}

\usepackage{multirow} %
\usepackage{tabularx} %
\usepackage{adjustbox} %
\usepackage{makecell} %
\usepackage[final]{microtype}

\usepackage{bm} %
\usepackage{mathtools} %

\usepackage{tabularx} 
\usepackage{amssymb}
\usepackage{amsmath,amsfonts}
\usepackage{algorithmic}
\usepackage{caption} 
\usepackage{textcomp}
\usepackage{xcolor}
\usepackage{todonotes}
\usepackage{xcolor}
\usepackage{comment}
\usepackage{tcolorbox}
\usepackage{colortbl} 
\usepackage{enumitem}
\usepackage{balance}
\usepackage{booktabs}

\usepackage{graphicx} %
\usepackage{array}    %
\usepackage{multirow} %

\begin{document}

\title{Evaluating DAO Sustainability and Longevity Through On-Chain Governance Metrics}

\author{Silvio Meneguzzo\thanks{Manuscript created March, 2025; This work was developed by Silvio Meneguzzo (Corresponding author) is with the Department of Computer Science, University of Turin, Italy (e-mail: silvio.meneguzzo@unito.it).}%
, Claudio Schifanella\thanks{Claudio Schifanella is with the Department of Informatics, University of Turin, Italy (e-mail: claudio.schifanella@unito.it).}%
, Valentina Gatteschi\thanks{Valentina Gatteschi is with the Department of Automation and Computer Science, Politecnico di Torino, Italy (e-mail: valentina.gatteschi@polito.it).}%
, Giuseppe Destefanis\thanks{Giuseppe Destefanis is with the Department of Computer Science, Brunel University of London, United Kingdom (e-mail: giuseppe.destefanis@brunel.ac.uk).}%
}

\maketitle


\begin{abstract}
Decentralised Autonomous Organisations (DAOs) automate governance and resource allocation through smart contracts, aiming to shift decision-making to distributed token holders. However, many DAOs face sustainability challenges linked to limited user participation, concentrated voting power, and technical design constraints. This paper addresses these issues by identifying research gaps in DAO evaluation and introducing a framework of Key Performance Indicators (KPIs) that capture governance efficiency, financial robustness, decentralisation, and community engagement. We apply the framework to a custom-built dataset of real-world DAOs constructed from on-chain data and analysed using non-parametric methods. The results reveal recurring governance patterns, including low participation rates and high proposer concentration, which may undermine long-term viability. The proposed KPIs offer a replicable, data-driven method for assessing DAO governance structures and identifying potential areas for improvement. These findings support a multidimensional approach to evaluating decentralised systems and provide practical tools for researchers and practitioners working to improve the resilience and effectiveness of DAO-based governance models.
\end{abstract}

\begin{IEEEkeywords}
Decentralized Autonomous Organizations, DAO, Blockchain, Voting Mechanisms, Decentralization, Governance, Sustainability, Longevity, Key Performance Indicators, User Participation.
\end{IEEEkeywords}

\section{Introduction}

\IEEEPARstart{D}{ecentralised} Autonomous Organisations (DAOs) introduce a governance model that replaces centralised decision-making with blockchain-based smart contracts and voting mechanisms \cite{buterin2014daos}. DAOs enable collective decision-making by users without the need for central authorities or intermediaries \cite{hassan2021decentralized}. This structure is based on decentralisation, transparency, and automated decision-making, making DAOs applicable to various collaborative systems.
Despite these characteristics, unresolved challenges remain \cite{tan2024openproblemsdaos}, particularly low user participation in governance \cite{feichtinger2023hidden}. When participants do not vote or engage in the decision-making process, a DAO’s ability to function effectively and maintain its decentralised structure is weakened \cite{HAN2025102734}. Addressing participation issues is necessary to ensure that DAOs operate as intended and support long-term sustainability.

In this paper, we examine how DAOs compare to traditional organisational models and analyse the blockchain-based mechanisms that influence participation and governance. By evaluating real-world DAOs through Key Performance Indicators (KPIs), we identify critical governance challenges related to user engagement and decision-making.

To address these issues, we define the following research questions:\\
\textbf{RQ1}: Which challenges most significantly affect DAO sustainability and longevity, particularly regarding user participation?\\
\textit{Rationale}: DAOs face multiple obstacles, but inadequate engagement is consistently identified as a primary factor influencing decentralisation and effectiveness.\\
\textbf{RQ2}: Which Key Performance Indicators (KPIs) can be used to evaluate DAO sustainability, including financial stability, governance processes, decentralisation, and community engagement?\\
\textit{Rationale}: Establishing measurable indicators allows for comparability and helps identify organisational and technical weaknesses in DAOs.\\
\textbf{RQ3}: How does applying these KPIs to real-world DAOs reveal governance issues and inform strategies for improving sustainability and longevity?\\
\textit{Rationale}: Beyond theoretical measures, the study aims to demonstrate how KPI-based analysis provides structured recommendations for improving DAO governance.

We make three principal contributions. First (I), we identify persistent open challenges in DAO governance, focusing specifically on the lack of consistent user engagement as a limiting factor for sustainability and effective decentralization. Second (II), we develop a set of empirically grounded KPIs spanning social, economic, and procedural dimensions of DAO governance. These KPIs offer a replicable and structured means of assessing DAOs across multiple dimensions, including voting activity, treasury management, automation, and distribution of power. Third (III), we apply these KPIs to a curated dataset based on on-chain data, comprising 50 active DAOs and demonstrate how the ensuing analysis reveals critical governance asymmetries, especially low voter turnout and concentration of proposer authority, and highlights paths for structural improvements. By establishing this integrated KPI framework, we aim to support both researchers and practitioners in diagnosing governance shortfalls and designing more resilient, community-driven DAO ecosystems.

By applying the proposed KPI framework to real-world cases, we show how inclusive governance structures, more equitable resource allocation, and user-friendly voting mechanisms can significantly boost engagement and decentralisation. In this sense, our findings suggest that structured, data-driven indicators enable DAOs to develop more participatory and automated governance models, ultimately strengthening their sustainability.
The rest of this paper is structured as follows: Section \ref{sec:background} reviews related work on DAOs, governance mechanisms, and ongoing challenges. Section \ref{sec:methodology} describes the research methodology, including KPIs development and the creation of a Multi-chain on-chain retrieval pipeline. Section \ref{sec:results} presents the results based on the analysis of existing DAOs. Section \ref{sec:discussion} discusses the implications of these findings, followed by threats to validity in Section \ref{sec:threats_validity}. Finally, Section \ref{sec:conclusion} provides conclusions and directions for future research.

We provide a replication package including all the nalysis scripts and results at this \href{https://figshare.com/s/2cf646c67f23ea917ac1}{link}\footnote{The replication package is hosted on Figshare: \url{https://figshare.com/s/2cf646c67f23ea917ac1}} to support
reproducibility and verification.

\section{Background}
\label{sec:background}

\subsection{DAOs: Definition and Evolution}

Decentralised Autonomous Organisations represent a governance model that uses blockchain technology and smart contracts to automate decision-making processes \cite{hassan2021decentralized}. Unlike traditional organisations, which rely on centralised authorities, DAOs distribute decision-making power among members through token-based voting systems \cite{wang2019decentralized}.

\noindent
The concept of DAOs emerged with Buterin’s \cite{buterin2014ethereum, buterin2014daos} introduction of Ethereum, the first blockchain platform supporting smart contracts capable of encoding organisational rules. In 2016, “The DAO” was launched as the first large-scale implementation, raising over \$150 million before a vulnerability in its smart contract led to a major security breach \cite{dupont2017experiments}. This event demonstrated how flaws in smart contract code could result in governance failures \cite{liu2021overview}. Interest in DAOs was later revived when MakerDAO introduced an on-chain governance system in 2018, followed by the adoption of similar approaches across various DeFi (Decentralized Finance) protocols \cite{altaleb2022decentralized}.

\subsection{DAO Characteristics and Challenges}

DAOs are built on three core principles: decentralisation (distributed network-based management), automation (code-based governance), and organisation (transparent operating rules via smart contracts) \cite{wang2023first}. Additional attributes include transparency, immutability, resistance to manipulation, interoperability, incentives for participation, and operational efficiency \cite{hassan2021decentralized, wright2020rise}.

Despite these characteristics, DAOs encounter significant challenges. Security vulnerabilities in smart contracts can have severe consequences due to blockchain immutability, as seen in “The DAO” hack \cite{wang2019decentralized}. Privacy concerns arise from the full transparency of all transactions \cite{liu2021overview}. Legal uncertainties persist in many jurisdictions, raising questions about liability \cite{wright2020rise}, despite the fact that some jurisdictions have begun formalizing DAO-friendly statutes, including Vermont \cite{vermont2018blockchain}, Wyoming \cite{wyoming2021dao}, the Marshall Islands \cite{marshallislands2022dao}, and Utah \cite{utah2023dao}. Technical constraints include the gap between legal frameworks and smart contract code \cite{wang2019decentralized}. Governance-related difficulties also remain, particularly the concentration of voting power and low participation rates \cite{feichtinger2023hidden} and related security issues in the governance process \cite{10891888}. Early analyses stressed that DAOs must balance on-chain rules with off-chain social coordination \cite{liu2019from}, but this approach is in contradiction with a full transparent and decentralized desired behaviour.

\subsection{DAO Governance Mechanisms}

Governance is a central aspect of DAOs, with voting mechanisms forming a key part of their structure alongside blockchain and smart contracts. Fan et al. \cite{fan2023insight} examined several voting mechanisms used in DAOs, including Approved Relative Majority, Token-Based Quorum, Quadratic Voting, Liquid Democracy, Weighted Voting, Rage Quitting, and Holographic Consensus. Ding et al. \cite{ding2023voting} also described Conviction Voting, which adjusts vote weight based on preference and time. Dimitri \cite{dimitri2023voting} highlights how different voting schemes, such as approval voting or rank-based voting, impact outcome legitimacy in DAO governance. Beck et al. \cite{beck2018governance} propose a blockchain governance model highlighting the tension between decentralization and the need for effective coordination.

\noindent
Despite the range of governance models, Feichtinger et al. \cite{feichtinger2023hidden} found that voting power remains concentrated in most DAOs. In their study of 21 governance systems, 17 were controlled by fewer than 10 participants. Common measures of voting power distribution include the Gini Coefficient and Nakamoto Coefficient \cite{fritsch2024analyzing}, which assess inequality and control concentration, respectively.
While these mechanisms represent a diverse range of voting protocols, prior empirical studies often overlook the interplay between on-chain user participation, treasury management, and the degree of proposal automation. As a result, comprehensive frameworks for capturing overall DAO sustainability remain underdeveloped. This gap underpins the need for an integrated KPI-based approach, as we discuss in Section \ref{sec:methodology}.

\subsection{Evaluation Approaches for DAOs}
Existing research on DAO evaluation often focuses on individual aspects rather than complete frameworks. Park et al. \cite{park2022toward} introduced dimensions for assessing decentralisation, while Faqir-Rhazoui et al. \cite{faqir2021comparative} conducted a comparative study of DAO platforms on Ethereum. Although these studies provide useful insights, they do not offer an integrated method for assessing sustainability.
Wang et al. \cite{wang2019decentralized} and Qin et al. \cite{qin2022web3} proposed architectural models to examine DAO structures, covering technological, execution, coordination, organisational, and application layers. These models provide a conceptual understanding of governance structures but do not translate into practical evaluation metrics.

\subsection{Research Gap}
Despite the growing body of literature on DAOs, there remains no widely accepted framework for comprehensively assessing their sustainability and longevity. Many studies rely on off-chain or aggregator data that fail to capture on-chain details and some of the existing works tend to focus narrowly on specific issues such as security vulnerabilities \cite{dupont2017experiments}, governance limitations \cite{feichtinger2023hidden}, or technical implementations \cite{wang2019decentralized}, while neglecting broader social and economic dimensions. For example, Park et al.\ \cite{park2022toward} propose decentralisation indicators focused on organisational structure, and Faqir-Rhazoui et al.\ \cite{faqir2021comparative} compare DAO platforms (Aragon, DAOstack, and DAOhaus) primarily in terms of feature sets. Although such studies contribute valuable insights into decentralisation ratios, treasury sizes, or code security, they rarely address procedural and community-oriented factors such as user participation, voting fairness, and long-term engagement within a unified evaluative model.

\noindent
Furthermore, the lack of a complete dataset incorporating detailed on-chain governance events further hampers progress. Many existing datasets offer incomplete views of token distribution and voting activities. A recent empirical study based on off-chain data from the Snapshot platform \cite{10891558} analysed 581 DAO projects over more than three years, providing valuable insights into DAO performance at scale. However, while such approaches capture broad activity metrics, they do not include the fine-grained, verifiable on-chain details, such as precise token transfers and smart contract event logs, needed for an integrated assessment of governance performance. In response, our work introduces a dataset, constructed directly from raw blockchain data, which enables a more accurate and integrated evaluation of DAO performance.

\noindent
Many DAOs suffer from persistently low voting turnout, centralised token ownership, or ambiguous legal standing \cite{liu2021overview, wright2020rise}, but a thorough, empirical framework that captures these varied challenges is still lacking. In particular, while indicators like the Gini Coefficient \cite{fritsch2024analyzing} or token-based voting measures \cite{fan2023insight} help quantify power concentration, they do not account for social incentives, governance processes, or community resilience; factors that are equally critical to a DAO’s viability over time.

\noindent
To address these gaps, this paper introduces a KPI framework to evaluate DAOs across four dimensions—participation, financial stability, voting efficiency, and decentralisation—linking social, economic, and procedural factors to identify governance issues and support long-term sustainability.

\section{Methodology}
\label{sec:methodology}

We used a data-based approach to develop and apply Key Performance Indicators for evaluating the sustainability and longevity of DAOs. Our methodology combined quantitative data analysis with qualitative reasoning to ensure that our KPIs were both empirically grounded and conceptually sound. Existing studies often focus on isolated aspects of DAO performance (e.g., decentralisation ratios, treasury sizes, or code security) and rely on incomplete aggregator-based datasets, which limits analyses of voting power distribution and user engagement  \cite{feichtinger2023hidden}, \cite{wang2019decentralized},
\cite{dupont2017experiments},
\cite{wang2023first}, \cite{10411467}.

\noindent
To address these limitations, we extracted raw data from multiple sources, including Smart Contracts' ABI from blockchain explorers (e.g., Etherscan) and direct queries to blockchain networks via providers like Infura and Alchemy. This process provided detailed numerical and categorical information on various aspects of DAOs, allowing reliable quantitative assessment of their structures and activities.

\noindent
We interpreted our quantitative findings within the context of existing research on DAO governance models, established theoretical frameworks, and current organisational patterns \cite{Rikken02102023}, \cite{10217072}. This combination ensured our KPIs captured both measurable indicators and broader aspects of DAO sustainability.

\subsection{Key Performance Indicators}

To assess the sustainability and longevity of DAOs, we defined four Key Performance Indicators covering core aspects such as community participation, financial capacity, governance processes, and decentralisation. These KPIs were shaped by existing literature \cite{faqir2021comparative, park2022toward, fan2023insight, feichtinger2023hidden, voterapathy} and insights from our curated dataset. Together, they offer a structured way to evaluate organisational resilience over time.

\subsubsection{KPI 1: Network Participation}

This KPI reflects the extent of member engagement within the DAO. Participation in voting and proposal creation is essential to decentralised governance and helps ensure representative decision-making, adaptability and helps mitigate oligarchic tendencies \cite{feichtinger2023hidden, voterapathy}. We define the Participation Rate as:

\begin{equation}
\label{eq1}
    \text{Participation Rate} = \left( \frac{\text{Active Members}}{\text{Total Members}} \right)
\end{equation}

An \emph{Active Member} is any address that has cast a vote or created a proposal. \emph{Total Members} are token holders with on-chain voting rights. Drawing on literature reporting low participation levels, we adopt the following classification:

\begin{itemize}
    \item \textbf{Low}: $<$10\%
    \item \textbf{Medium}: 10--40\%
    \item \textbf{High}: $>$40\%
\end{itemize}

\noindent
These categories reflect empirical observations of voter turnout in prominent DAOs \cite{feichtinger2023hidden} and distinguish between minimal, moderate, and widespread participation. Prior analyses indicate that below 10\% turnout, governance tends to be dominated by a handful of token holders, undermining true decentralization of partecipation.

\subsubsection{KPI 2: Accumulated Funds}
\label{sec:kpi_funds}

This KPI captures a DAO’s financial capacity. We consider two aspects:

\noindent
\textbf{Treasury Size} denotes the total assets held in smart contracts under DAO control. Larger treasuries allow for sustained contributor rewards, protocol development, and other activities.

\noindent
\textbf{Circulating Token Percentage} measures the proportion of governance tokens in circulation:

\begin{equation}
\label{eq2}
    \text{Circulating Token Percentage} = \left( \frac{\text{Circulating Supply}}{\text{Total Supply}} \right)
\end{equation}

\noindent
Tokens held by the treasury or locked in vesting contracts are excluded from circulation. High circulation suggests broader economic participation, while low circulation may indicate concentration. We combine both aspects to categorise financial status:

\begin{itemize}
    \item \textbf{Low}: Treasury $<$\$100M
    \item \textbf{Medium-Low}: \$100M--\$1B, circulation $\leq$ 50\%
    \item \textbf{Medium-High}: \$100M--\$1B, circulation $>$ 50\%
    \item \textbf{High}: Treasury $>$\$1B
\end{itemize}

\noindent
Treasury size thresholds at \$100 million and \$1 billion reflect typical boundaries observed in leading DeFi DAOs like MakerDAO, Compound, and Uniswap, where financial resources can significantly influence governance dynamics and system resilience.

\subsubsection{KPI 3: Voting Mechanism Efficiency}

This KPI considers governance effectiveness based on proposal approval rate and voting duration:

\begin{equation}
\label{eq3}
    \text{Approval Rate} = \left( \frac{\text{Approved Proposals}}{\text{Total Proposals}} \right)
\end{equation}

\begin{equation}
\label{eq4}
    \text{Average Voting Duration} = \frac{\sum_{i=1}^{n} \text{Voting Duration}_i}{n}
\end{equation}

\noindent
Short durations may indicate rushed decision-making, while longer voting windows can hinder timely execution. Based on empirical observations \cite{fan2023insight}, we define:

\begin{itemize}
    \item \textbf{Low}: Approval $<$30\% and/or duration $<$3 days
    \item \textbf{Medium}: Approval 30--70\%, duration 3--14 days
    \item \textbf{High}: Approval $>$70\%, duration 3--14 days
\end{itemize}

\noindent
These levels help distinguish between inefficient governance, functional deliberation, and overly streamlined approvals.

\subsubsection{KPI 4: Decentralisation}

This KPI addresses the concentration of resources and the degree of autonomy in operations. It combines token distribution, member activity, and automation. Inspired by work on DAO structures \cite{park2022toward}, \cite{LOMONACO2025112233}, we assess decentralisation based on the largest token holder’s share, whether there is sufficient participation, and whether decisions are executed automatically.
We label a DAO’s decisions as “fully automated” when successful on-chain proposals directly trigger contract execution without requiring off-chain signatures or multi-sig confirmations.
\begin{itemize}
    \item \textbf{Low}: Largest holder $>$66\%
    \item \textbf{Medium-Low}: 33--66\%
    \item \textbf{Medium}: 10--33\%, with at least medium participation, no automation
    \item \textbf{Medium-High}: 10--33\%, with medium/high participation, full automation
    \item \textbf{High}: $<$10\%
\end{itemize}

\noindent
Our on-chain dataset allows precise measurement of token distributions and automation status. These categories differentiate between DAOs with strong individual control and those exhibiting distributed ownership and autonomous operations.

\subsubsection{Scoring System}

Each KPI level is mapped to a numeric score from 0 to 3, with all four KPIs equally weighted. The total score ranges from 0 to 12. This uniform scheme avoids arbitrary prioritisation and reflects our assumption that DAO sustainability results from balanced performance across community, financial, procedural, and structural domains. Table~\ref{table:kpi_levels} summarises the scoring.

\begin{table}[h!]
\centering
\caption{KPIs and Their Level Divisions with Assigned Scores}
\label{table:kpi_levels}
\resizebox{\columnwidth}{!}{
\begin{tabular}{| c || c | m{7cm} | c |}
 \hline
 \textbf{KPI} & \textbf{Level} & \textbf{Description} & \textbf{Score} \\ 
 \hline\hline
 \multirow{3}{*}{Network Participation} 
 & Low & Participation rate $<$ 10\% & 1 \\
 & Medium & Participation rate 11\%--40\% & 2 \\
 & High & Participation rate $>$ 40\% & 3 \\
 \hline
 \multirow{4}{*}{Accumulated Funds} 
 & Low & Treasury $<$ \$100 million USD & 0.75 \\ 
 & Medium-Low & Treasury \$100 million--\$1 billion USD, circulating tokens $\leq$ 50\% & 1.5 \\
 & Medium-High & Treasury \$100 million--\$1 billion USD, circulating tokens $>$ 50\% & 2.25 \\
 & High & Treasury $>$ \$1 billion USD & 3 \\
 \hline
 \multirow{3}{*}{Voting Mechanism Efficiency} 
 & Low & Approval rate $<$ 30\% and/or voting duration $<$ 3 days & 1 \\ 
 & Medium & Approval rate 30\%--70\%, voting duration 3--14 days & 2 \\
 & High & Approval rate $>$ 70\%, voting duration 3--14 days & 3 \\
 \hline
 \multirow{5}{*}{Decentralisation} 
 & Low & Largest holder $\geq$ 66\% of resources & 0.6 \\ 
 & Medium-Low & Largest holder 33\%--66\% of resources & 1.2 \\
 & Medium & Largest holder 10\%--33\%, medium participation, not fully automated decisions & 1.8 \\
 & Medium-High & Largest holder 10\%--33\%, medium/high participation, fully automated decisions & 2.4 \\
 & High & Largest holder $<$ 10\% of resources & 3 \\
 \hline
\end{tabular}
}
\end{table}

\noindent
This scoring method reflected our assumption that DAO sustainability depends on a balanced combination of social (participation), financial (funds), procedural (voting efficiency), and organisational (decentralisation) factors. We used equal weights to avoid introducing arbitrary prioritisation, though future work may adjust this based on further analysis.

\noindent
By combining these four KPIs, each based on our harmonised on-chain dataset, we developed a multidimensional view of DAO sustainability and enabled comparative analysis across different governance structures.

\subsection{Dataset Construction}

Our initial data collection included 5,999 DAOs from aggregator sources, most notably, the DAO Analyzer dataset from Kaggle. However, after applying strict criteria for on-chain voting and recent governance activity, most entries failed to meet our thresholds for meaningful participation\footnote{\url{https://github.com/smeneguz/data-analyzer-dao-ecosystem.git}}.In particular, only a negligible number of DAOs (2 DAOs) from the Kaggle dataset exhibited the required level of governance activity; hence, we opted to exclude this dataset from our final analysis. To address this, we developed the custom pipeline described in Section~\ref{subsubsec:onchain_pipeline}, creating a harmonized on-chain dataset comprising 50 DAOs that demonstrated:

\begin{itemize}
    \item Robust on-chain governance mechanisms
    \item Transparent participation profiles
    \item Key activity metrics (e.g., average voting duration, proposal frequency, treasury size)
\end{itemize}

\noindent
We classified DAOs into four activity categories:

\begin{enumerate}
    \item \textbf{Highly Active}: DAOs with at least 5 governance-related transactions in the last 30 days, showing consistent involvement from multiple members.
    \item \textbf{Moderately Active}: DAOs with at least 1 transaction/proposal in the last 90 days, maintaining regular community activity.
    \item \textbf{Minimally Active}: DAOs with transactions/proposals older than 90 days and low recent activity.
    \item \textbf{Potential Test or Dormant}: DAOs with fewer than 2 transactions since creation, possibly representing experimental deployments or abandoned projects.
\end{enumerate}

\noindent
These thresholds follow conventions in blockchain governance literature, which often uses monthly and quarterly assessments to track participation trends \cite{faqir2021comparative}.

Before calculating KPIs, we performed several validation checks:
\begin{itemize}
    \item Removal of duplicates
    \item Consistency checks across sources
    \item Standardisation of timestamps
    \item Verification of event logs by comparing proposal creation and execution events
\end{itemize}

\noindent
Any inconsistencies prompted targeted queries to blockchain nodes or re-examination of aggregator data, helping ensure a reliable dataset.

\subsection{Data Collection Pipeline}
\label{subsubsec:onchain_pipeline}

\begin{figure*}[!t]
    \centering
    \includegraphics[width=\textwidth]{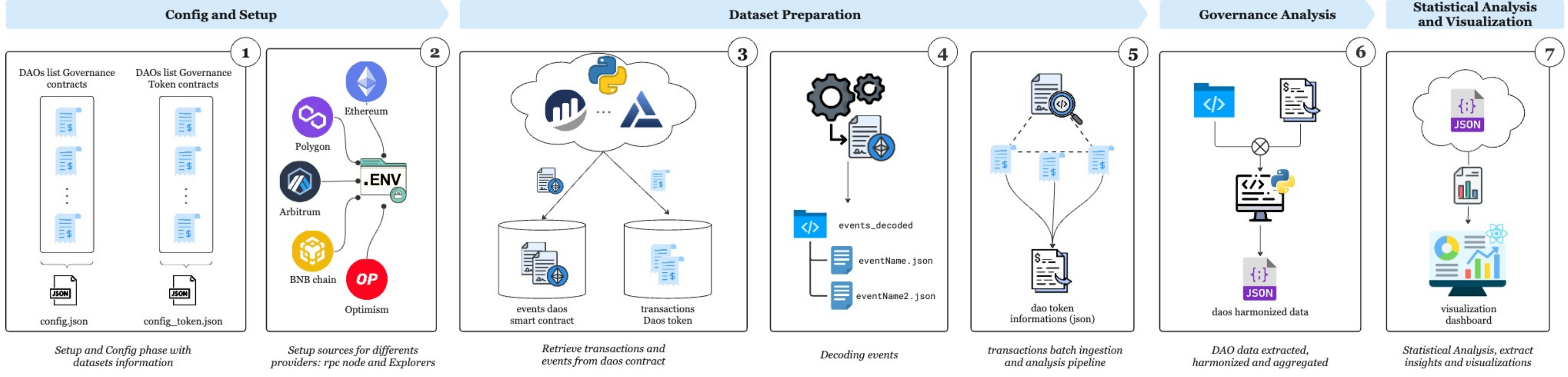}
    \caption{Multi-chain on-chain retrieval pipeline}
    \label{fig:pipeline_diagram}
\end{figure*}

Understanding DAO dynamics required thorough data extraction and analysis from multiple sources. Our multi-chain on-chain retrieval pipeline, illustrated in Figure~\ref{fig:pipeline_diagram}, consisted of the following steps:\\
\textbf{DAO and Token Address Enumeration} (Step 1 in Figure~\ref{fig:pipeline_diagram}): We identified candidate DAOs on Ethereum and other EVM-compatible networks (e.g., Polygon, Arbitrum, BNB Chain) by scanning known governance contracts and verifying that they supported on-chain voting transactions. 

\noindent
As shown in the leftmost panels of Figure~\ref{fig:pipeline_diagram}, we processed governance contracts and token contracts separately, organizing their addresses and interfaces into config files for subsequent processing.\\
\textbf{Setup for Multi-chain Access} (Step 2 in Figure~\ref{fig:pipeline_diagram}): We configured connections to multiple blockchain networks through node providers and explorers. As shown in the second panel of Figure~\ref{fig:pipeline_diagram}, this involved setting up environment configurations (.ENV files) to access Ethereum, Polygon, Arbitrum, BNB Chain, and Optimism networks, ensuring broad coverage of DAO activity.\\
\textbf{Smart Contract Event Retrieval} (Step 3 in Figure~\ref{fig:pipeline_diagram}): For each DAO's governance contract(s) and associated token(s), we queried block explorers and node providers to fetch all event logs from the deployment block until April 2025. In parallel, we retrieved transactions from the associated governance token smart contracts to capture token transfers and other economic events. As shown in the third panel of Figure~\ref{fig:pipeline_diagram}, this process created separate datasets for smart contract events and token transactions, which were stored in dedicated databases for later analysis.\\
\textbf{Event Decoding and Data Normalisation} (Step 4 in Figure~\ref{fig:pipeline_diagram}): We decoded each raw log using the contract's ABI to map event signatures into human-readable records. We then converted timestamps to UTC, normalized token amounts, and filtered out repeated or malformed entries. The fourth panel of Figure~\ref{fig:pipeline_diagram} illustrates how we transformed raw blockchain data into structured event files (events\_decoded, eventName.json, eventName2.json), making them suitable for analysis.\\
\textbf{Cross-linking Governance and Token Data} (Step 5 in Figure~\ref{fig:pipeline_diagram}): To assess how voting power correlated with participation, we supplemented event data with historical token transfers retrieved via RPC nodes, enabling us to track token holder distribution over time and identify concentrated voting power. As shown in the fifth panel of Figure~\ref{fig:pipeline_diagram}, we processed token information into standardized JSON formats that captured ownership patterns and voting activity.\\
\textbf{DAO Data Extraction and Analysis} (Step 6 in Figure~\ref{fig:pipeline_diagram}): We integrated governance events with token distribution data to create a consistent view of each DAO's activity. The sixth panel of Figure~\ref{fig:pipeline_diagram} shows how we combined these datasets into a unified format that enabled consistent analysis across different DAOs and governance models.\\
\textbf{Harmonised Dataset Output and Visualization} (Step 7 in Figure~\ref{fig:pipeline_diagram}): Finally, we generated standardized files for each DAO, including key fields such as total proposals, voter addresses, top holders, and execution outcomes. This structured output, represented in the rightmost panel of Figure~\ref{fig:pipeline_diagram}, formed the basis for KPI calculation, statistical analysis, and visualization through customised dashboards.

\section{Results}
\label{sec:results}
Our sample consists of 50 DAOs spread across Ethereum, Polygon, and Arbitrum. Together, they account for 6930 proposals, 317317 unique voting addresses and 4524205 total members, providing a robust basis for evaluating governance patterns.
We applied the Shapiro--Wilk test for normality and Levene's test for variance homogeneity to assess the suitability of parametric methods. Based on the results of these tests, we used one-way ANOVA for normally distributed groups with homogeneous variances, and the Kruskal--Wallis test for cases where these assumptions were violated. To complement the statistical analysis, we generated box plots, violin plots, scatter plots with regression lines and radar charts. These visualisations illustrate the relationships among the KPI metrics and facilitate the empirical evaluation of the proposed framework using our newly compiled dataset of 50 DAOs.

\subsection{Statistical Approach}
\label{sec:stat_approach}

We structured the data analysis as a sequence of standard tests to verify statistical assumptions and ensure that group comparisons were conducted appropriately.

\subsubsection{Shapiro--Wilk Test for Normality}

For each KPI category (e.g.\ \emph{Low}, \emph{Medium}, \emph{High}), we first evaluated whether the data followed a normal distribution. The Shapiro--Wilk test \cite{eb32428d-e089-3d0c-8541-5f3e8f273532} was used, with the test statistic $W$ computed as:

\[
W \;=\; \frac{\Bigl(\sum_{i=1}^{n} a_i\, x_{(i)}\Bigr)^{2}}
{\sum_{i=1}^{n} (x_i - \bar{x})^{2}},
\]

where $x_{(i)}$ denotes the $i$-th order statistic (i.e., the sample sorted in ascending order), $\bar{x}$ is the sample mean, and $a_i$ are constants derived from the covariance matrix of a normal distribution. A \emph{p}-value below a commonly used threshold (0.05) led to rejection of the null hypothesis of normality. Given the Shapiro--Wilk test’s sensitivity to small sample sizes, we report its results only for categories with $n \ge 3$.

\subsubsection{Levene’s Test for Variance Homogeneity}

We used Levene’s test \cite{Levene1961RobustTF} to assess whether variances were homogeneous across groups. For $k$ groups, let $X_{ij}$ denote the $j$-th observation in the $i$-th group, with $i = 1,\dots,k$. The test statistic is computed as:

\[
W_{\mathrm{Levene}} \;=\;
\dfrac{(N-k)\sum_{i=1}^k n_i \bigl(Z_{i\cdot}-Z_{\cdot\cdot}\bigr)^2}
{(k-1)\sum_{i=1}^k \sum_{j=1}^{n_i} (Z_{ij}-Z_{i\cdot})^2},
\]

where $Z_{ij} = \lvert X_{ij} - \bar{X}_i \rvert$ (or, alternatively, the median may be used in place of the mean), $Z_{i\cdot}$ is the mean of group $i$ after transformation, and $Z_{\cdot\cdot}$ is the overall mean of all $Z_{ij}$. $N$ denotes the total number of observations, and $n_i$ is the size of group $i$. A significant \emph{p}-value (typically $<0.05$) indicates heterogeneity of variances across groups.

\subsubsection{Parametric vs.\ Non-parametric Group Comparisons}

\textbf{(1) One-Way ANOVA:} If all groups satisfied the normality assumption (Shapiro--Wilk) and exhibited homogeneous variances (Levene’s), we applied one-way Analysis of Variance (ANOVA) to test for differences in group means. The ANOVA $F$-statistic was evaluated against the null hypothesis that all group means are equal \cite{STHLE1989259}.

\textbf{(2) Kruskal--Wallis Test:} When either normality or homogeneity of variances was not satisfied, we used the Kruskal--Wallis test \cite{ac6c544c-0197-38bd-8c06-ec4f655ff4fd}:

\[
H \;=\;
\dfrac{12}{N(N+1)} \sum_{i=1}^{k} n_i\,R_i^{2}
\;-\; 3\,(N+1),
\]

where $N$ is the total sample size across $k$ groups, $n_i$ is the sample size of group $i$, and $R_i$ is the average rank within group $i$. The null hypothesis states that all groups follow the same distribution. A significant \emph{p}-value indicates that at least one group differs. Post-hoc pairwise comparisons were conducted using Dunn’s test with Bonferroni correction to identify specific group differences.

\subsubsection{Interpretation of Test Outcomes}

\noindent To guide the application of statistical tests, we established a decision process based on the outcomes of normality and variance homogeneity checks. Table~\ref{tab:test_outcomes} summarises the interpretation criteria for the Shapiro--Wilk, Levene’s, ANOVA, and Kruskal--Wallis tests, along with the corresponding implications for group comparison methods.

\begin{table}[h]
\centering
\caption{Interpretation of statistical test outcomes and corresponding analysis decisions.}
\resizebox{\columnwidth}{!}{
\begin{tabular}{@{}ll@{}}
\toprule
\textbf{Condition} & \textbf{Interpretation} \\
\midrule
Shapiro--Wilk \emph{p}-value $< 0.05$ & Data deviate from normality; ANOVA not used. \\
Levene’s Test \emph{p}-value $< 0.05$ & Variances differ; ANOVA not used. \\
ANOVA \emph{p}-value $< 0.05$ & At least one group mean differs significantly. \\
Kruskal--Wallis \emph{p}-value $< 0.05$ & At least one group distribution differs; post-hoc tests applied. \\
\bottomrule
\end{tabular}
}
\label{tab:test_outcomes}
\end{table}

\noindent This procedure ensures that each KPI category is analysed using appropriate statistical methods, reducing the risk of misinterpretation due to violations of parametric assumptions. The following sections present the results for each KPI, including the applied thresholds and visualisations.

\subsection{Network Participation}
\label{sec:res_kpiA}

\noindent
\textbf{Definition.} As introduced in Section~\ref{sec:methodology}, Network Participation measures the proportion of \emph{active} members, those who cast at least one on-chain vote or submitted a proposal, relative to total membership. We classified DAOs into three categories: \emph{Low} ($<10\%$), \emph{Medium} ($10\text{--}40\%$), and \emph{High} ($>40\%$).

\begin{figure}[!t]
\centering
\includegraphics[width=0.47\textwidth]{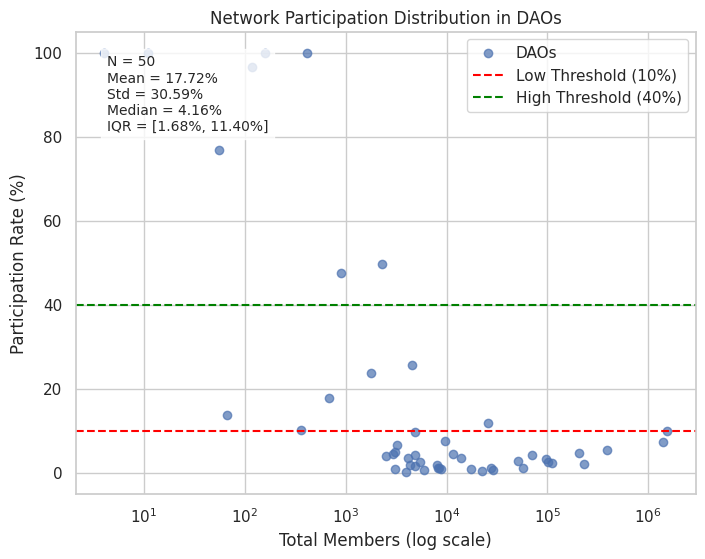}
\caption{Scatter plot showing the
relationship between total members (x) and participation
rate (y)}
\label{fig:scatter_participation}
\end{figure}

\noindent
\textbf{Findings.} Figure~\ref{fig:scatter_participation} presents the relationship between total membership (log-scaled $x$-axis) and participation rate ($y$-axis). The visualisation highlights an inverse pattern: smaller DAOs tend to show higher participation, with thresholds at 10\% and 40\% marking the category boundaries. Summary statistics indicate a low median participation rate (4.16\%) and high variability, with a few outlier values capped at 100\%

\begin{figure}[!t]
\centering
\includegraphics[width=0.47\textwidth]{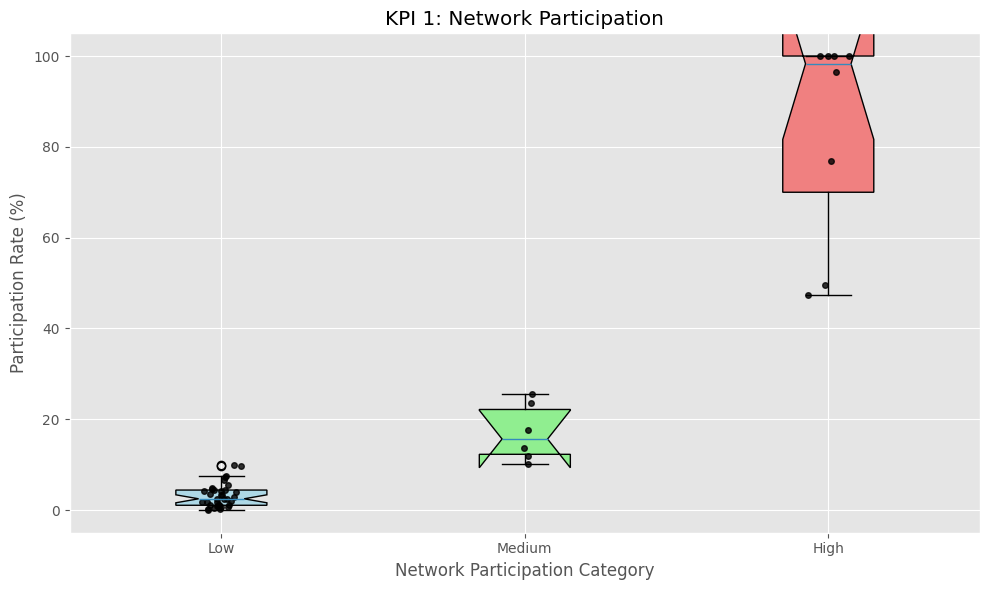}
\caption{"Notched`` box plot for Network Participation.}
\label{fig:participation_boxplot}
\end{figure}

\noindent
Figure~\ref{fig:participation_boxplot} shows notched box plots across the three participation categories. The notches represent 95\% confidence intervals for the medians. The Shapiro--Wilk test returned \emph{p}-values below 0.05 for the Low and High categories, and Levene’s test indicated unequal variances. Based on these results, we applied the Kruskal--Wallis test, which identified significant group differences ($H = 30.45$, $p < 0.01$). Post-hoc analysis using Dunn’s test with Bonferroni correction confirmed that the High group had a significantly higher median participation rate (median = 98.29\%) than the Low group (median = 2.47\%), confirming a substantial gap in user engagement across the DAO ecosystem.

\subsection{Accumulated Funds}

\noindent
\textbf{Definition.} Accumulated Funds reflect the DAO’s financial capacity, incorporating both treasury size and the proportion of circulating tokens. As defined in Section~\ref{sec:methodology}, DAOs were classified into four categories: \emph{Low}, \emph{Medium-Low}, \emph{Medium-High}, and \emph{High}, based on combined thresholds for these two dimensions.

\begin{figure}[!t]
\centering
\includegraphics[width=0.47\textwidth]{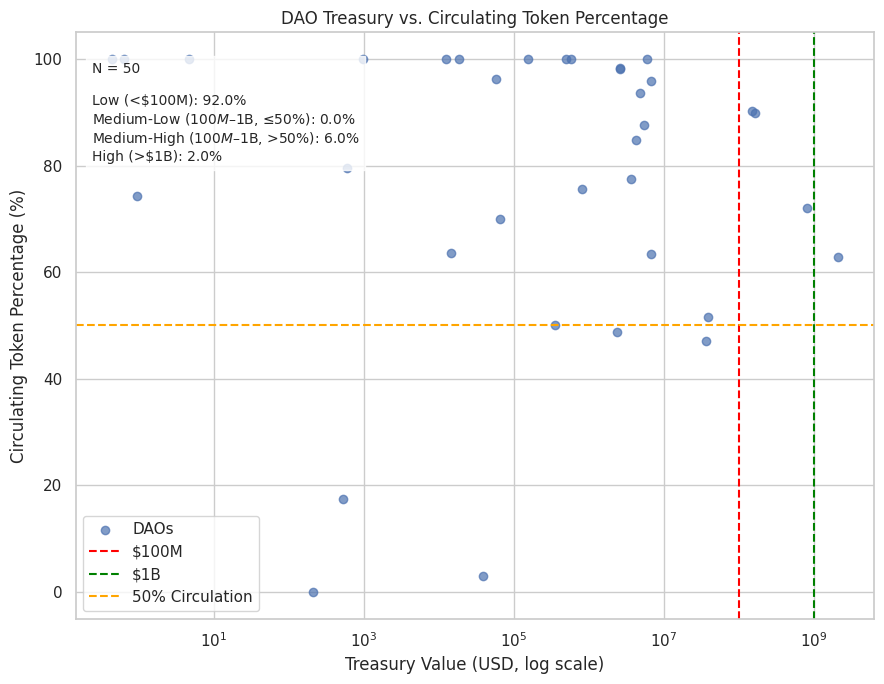}
\caption{Scatter plot of treasury value (log scale) vs.\ circulating token percentage.}
\label{fig:general_funds_scatter}
\end{figure}

\noindent
\textbf{Findings.} Figure~\ref{fig:general_funds_scatter} visualises the distribution of DAOs by plotting treasury value (in \textit{log scale} to handle wide differences from \$1 million to billions) against circulating token percentage. Threshold lines at \$100\,million, \$1\,billion, and 50\% token circulation delineate category boundaries.

\begin{figure}[!t]
\centering
\includegraphics[width=0.47\textwidth]{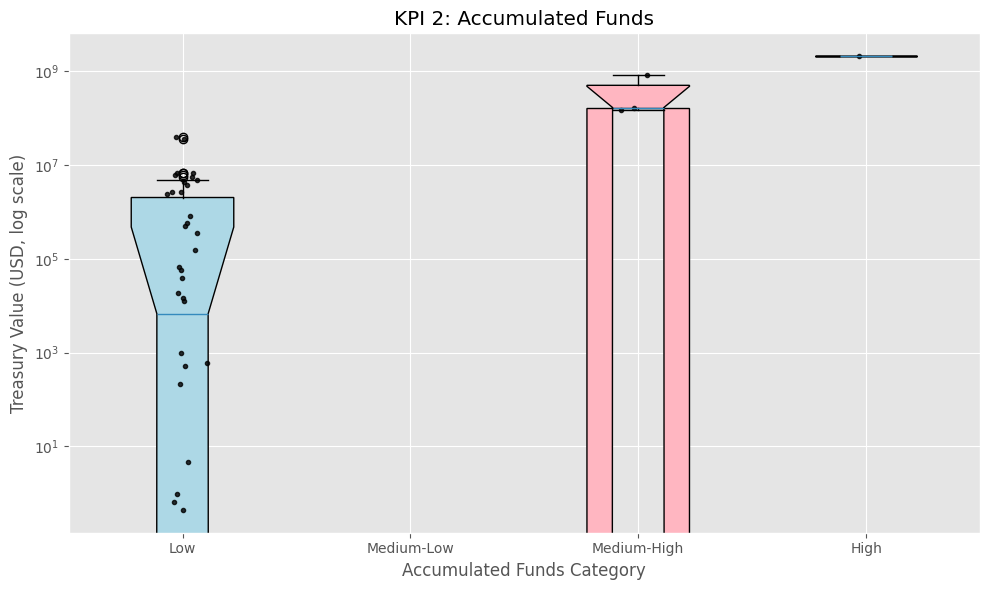}
\caption{Notched box plot of treasury sizes, grouped by circulating token status.}
\label{fig:funds_boxplot}
\end{figure}

\noindent
Figure~\ref{fig:funds_boxplot} shows notched box plots of treasury sizes grouped by circulating token status. DAOs in the \emph{Low} category (treasury $< \$100$ million) exhibited high variance, possibly reflecting early-stage operations or inconsistent funding cycles. Normality tests (Shapiro--Wilk) returned \emph{p}-values below 0.05 for all four categories, and Levene’s test indicated heterogeneity of variances. These results support the use of non-parametric methods for group comparisons.

\noindent
\textbf{Visualisation Support.} We observed a weak and statistically non-significant correlation between treasury size and circulating token percentage (Pearson’s $r \approx -0.13$, $p \approx 0.37$).

\subsection{Voting Mechanism Efficiency}
\label{sec:res_kpiC}

\noindent
\textbf{Definition.} Voting Mechanism Efficiency captures the balance between approval rates and average voting duration, grouping DAOs into \emph{Low}, \emph{Medium}, or \emph{High} efficiency categories. This KPI reflects the trade-off between swift decision-making and thorough deliberation.

\begin{figure}[!t]
\centering
\includegraphics[width=0.47\textwidth]{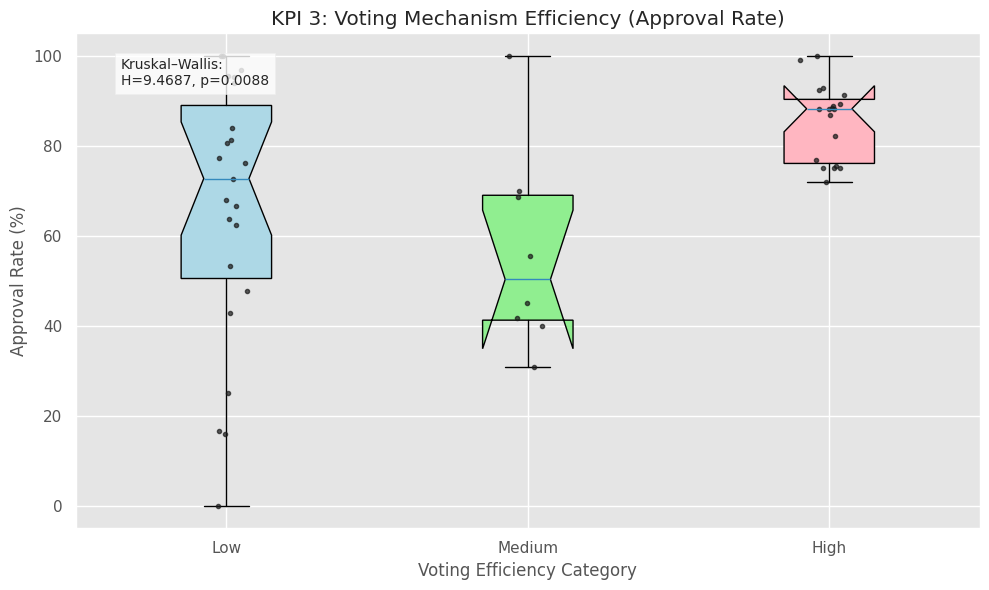}
\caption{Notched box plot of approval rates among the three efficiency categories}
\label{fig:voting_boxplot}
\end{figure}

\noindent
\textbf{Findings.} Figure~\ref{fig:voting_boxplot} presents notched box plots of approval rates across the three efficiency categories. The Shapiro--Wilk test indicated non-normality in the \emph{Low} ($p = 0.0373$) group, while the \emph{Medium} ($p = 0.4003$) and \emph{High} ($p = 0.0804$) groups satisfied the normality assumption. Levene’s test confirmed variance heterogeneity ($p = 0.0043$). Accordingly, we applied the Kruskal--Wallis test, which returned $H = 9.4687$ and $p = 0.0088$. Although the \emph{High} group showed the highest median approval rate (88.24\%), the differences were not statistically significant at the 5\% level.

\begin{figure}[!t]
\centering
\includegraphics[width=0.47\textwidth]{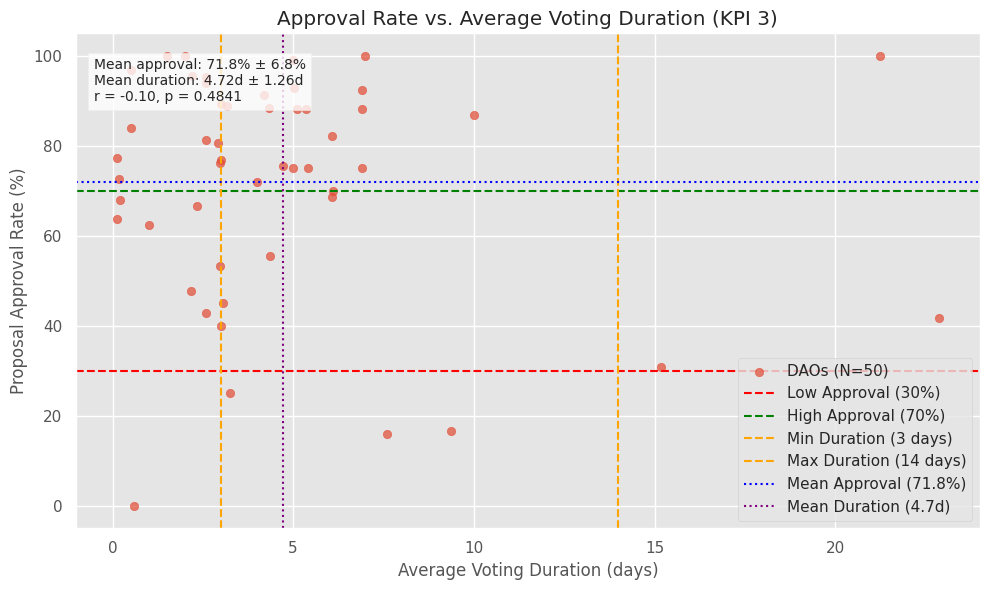}
\caption{Scatter plot of \emph{approval rate} versus \emph{average voting duration}}
\label{fig:voting_scatter}
\end{figure}

\noindent
\textbf{Visualisation Support.} Figure~\ref{fig:voting_scatter} displays approval rate against average voting duration, offering a joint view of the two dimensions that define this KPI. While no direct statistical inference is drawn, the pattern suggests that extremely short or excessively long voting windows may undermine effective governance. These empirical observations are consistent with prior findings \cite{fan2023insight} recommending moderate voting periods in DAO settings.

\subsection{Decentralisation}

\noindent
\textbf{Definition.} Decentralisation encompasses economic distribution, participatory engagement, and the degree of on-chain automation \cite{park2022toward}. We classified DAOs into five categories: \emph{Low}, \emph{Medium-Low}, \emph{Medium}, \emph{Medium-High}, and \emph{High},  based on the largest holder’s token share and the presence of automated governance processes.

\begin{figure}[!t]
\centering
\includegraphics[width=0.48\textwidth]{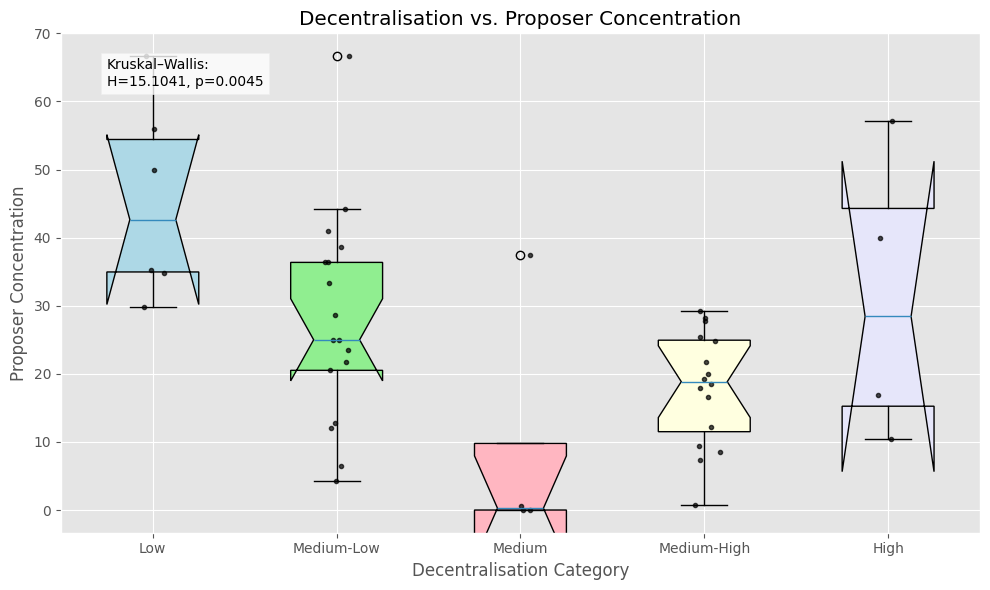}
\caption{Notched box plot of \emph{proposer concentration} across five decentralisation categories}
\label{fig:decentralisation_boxplot}
\end{figure}

\noindent
\textbf{Findings.} Figure~\ref{fig:decentralisation_boxplot} shows notched box plots of proposer concentration across the five decentralisation categories. The Shapiro--Wilk test indicated that the \emph{Medium} group ($p = 0.0018$) deviated from normality, while other groups did not violate normality assumptions. Levene’s test showed no significant variance heterogeneity ($p = 0.2338$). We therefore applied the Kruskal--Wallis test, which returned $H = 15.10$ and $p = 0.0045$, indicating statistically significant differences in proposer concentration among the decentralisation categories.
\noindent
The \emph{Low} decentralisation group (largest holder $>$ 66\%) exhibited the highest mean proposer concentration (45.41), while the \emph{Medium} group (10–33\% ownership, no automation) had the lowest mean (9.51). These results suggest that high token concentration does not necessarily limit proposal activity, whereas intermediate ownership levels may correspond to reduced proposer diversity in this dataset.

\begin{figure}[!t]
\centering
\includegraphics[width=0.47\textwidth]{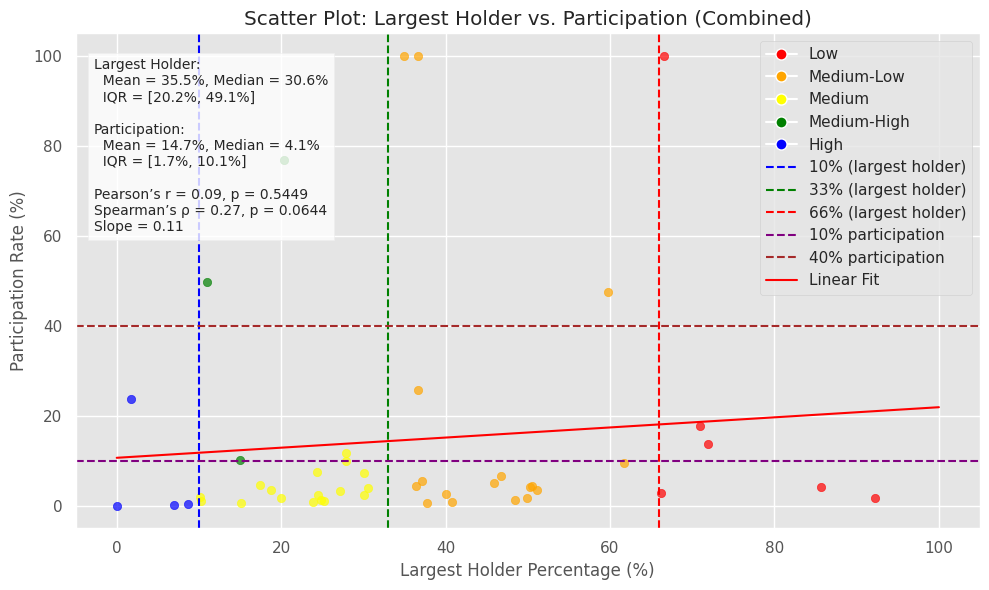}
\caption{Scatter plot of \emph{largest holder percentage} versus \emph{participation rate}
}
\label{fig:decentralisation_scatter}
\end{figure}

\noindent
\textbf{Visualisation Support.} Figure~\ref{fig:decentralisation_scatter} maps the largest holder’s percentage against participation rate. The Pearson correlation coefficient was weak and non-significant ($r = 0.09$, $p = 0.5449$), indicating no linear relationship. However, Spearman’s rank correlation showed a moderate monotonic trend ($\rho = 0.27$, $p = 0.0644$), suggesting a mild association between lower concentration and higher engagement. Threshold lines at 10\%, 33\%, and 66\% (economic decentralisation) and at 10\% and 40\% (participation) provide reference boundaries. While a slight rank-based trend is observable, the visual evidence does not indicate a strong predictive relationship between largest-holder share and participation rate.

\subsection{Composite Metrics and Overall Patterns}
\label{subsec:overall_visuals}

\noindent
To summarise DAO performance across dimensions, we constructed composite scores based on the four KPIs: \emph{Network Participation}, \emph{Accumulated Funds}, \emph{Voting Mechanism Efficiency}, and \emph{Decentralisation}.

\begin{figure}[!t]
\centering
\includegraphics[width=0.48\textwidth]{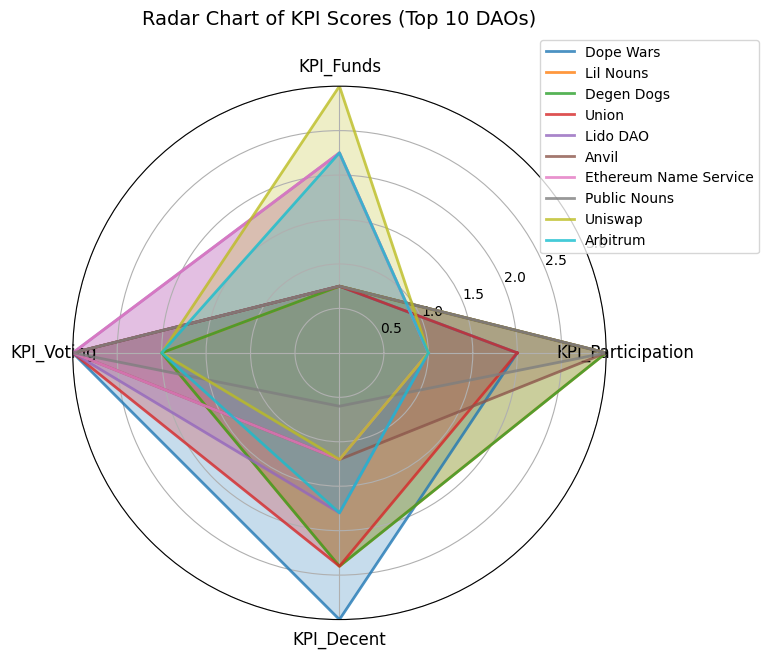}
\caption{radar composite plot - 10 daos}
\label{fig:radar_composite}
\end{figure}

\noindent
Figure~\ref{fig:radar_composite} presents radar plots for a subset of DAOs, illustrating the trade-offs and balance across the four governance dimensions. DAOs with more balanced profiles, showing consistent scores across all KPIs, tend to achieve higher composite scores, suggesting more resilient governance structures. In contrast, DAOs that perform strongly in one or two KPIs but poorly in others often display structural imbalances that may affect long-term sustainability.

\noindent
For example, Uniswap and Lido DAO demonstrate high financial capacity but score lower in decentralisation. By contrast, DAOs such as Public Nouns, Lil Nouns, and Union achieve high composite scores through more distributed governance structures and consistent performance across participation and voting dimensions. DAOs with lower composite scores, such as HAI, Open Dollar, and Unlock, often reflect a combination of centralised ownership and limited user engagement.

\noindent
These aggregated comparisons reinforce the core premise of the study: assessing DAO sustainability requires a multi-dimensional perspective. Strong performance in one area does not necessarily compensate for weaknesses in others. A combined evaluation of participation, financial health, procedural efficiency, and decentralisation provides a more complete understanding of organisational robustness.

\section{Discussion}
\label{sec:discussion}

\subsection{Interpretation of KPI Findings}
\label{subsec:discussion_kpifindings}

\noindent
The four KPIs, Network Participation, Accumulated Funds, Voting Mechanism Efficiency, and Decentralisation, yielded statistically significant group differences in most cases, based primarily on non-parametric testing. DAOs classified as \emph{High} in Network Participation and Accumulated Funds exhibited more consistent engagement and greater financial capacity, respectively. Those with Medium-High to High decentralisation levels showed broader proposer distributions and lower voting concentration. In contrast, correlations such as that between largest-holder percentage and participation were weaker or marginal, indicating that concentrated token ownership does not necessarily reduce member activity.

\noindent
DAOs scoring highly across multiple KPIs demonstrated structural features associated with more sustainable governance. These findings support the broader view that sustained community involvement, procedural efficiency, and equitable resource distribution jointly contribute to the resilience of decentralised organisations.

\subsection{Degree of Decentralisation}
\label{subsec:discussion_decentralisation}

\noindent
The results show that many DAOs exhibit partial decentralisation, with voting activity or proposal initiation often concentrated among a small number of addresses. However, several cases of high decentralisation were observed, particularly where moderate to large treasuries were combined with automated on-chain governance mechanisms. These patterns suggest that decentralisation is shaped not only by token distribution but also by social and procedural dynamics such as proposer diversity, governance automation, and participation incentives.

\noindent
Notably, variations in the largest-holder percentage did not consistently suppress network participation. This finding complicates assumptions about centralisation effects, suggesting that concentrated capital, when coupled with effective delegation or automation, may coexist with active governance.

\subsection{Voting Mechanism Efficiency}
\label{subsec:discussion_votingefficiency}

\noindent
While DAOs classified as \emph{High} in voting efficiency generally exhibited shorter but adequate voting durations and higher approval rates, not all observed differences across groups were statistically significant (Section~\ref{sec:res_kpiC}). This outcome points to the influence of contextual factors, such as proposal complexity or urgency, on governance behaviour. Shorter decision cycles do not inherently imply more effective outcomes.

\noindent
Scatter plots of approval rate versus average voting duration suggested that overly short windows may hinder deliberation, while excessively long ones can reduce engagement. A more dynamic approach, calibrating voting duration based on prior proposal complexity or participation history, could improve legitimacy without sacrificing efficiency. These observations suggest that future refinements to the boundaries for KPI 3 may strengthen its interpretive value.

\subsection{Implications for DAO Governance}
\label{subsec:discussion_implications}

\noindent
The KPI framework enables a structured diagnosis of DAO governance strengths and weaknesses. DAOs with limited participation but substantial financial reserves may benefit from changes to voting accessibility or community engagement strategies. Conversely, DAOs with strong participation but weaker financial capacity may need to diversify treasury structures or enhance economic sustainability mechanisms.

\noindent
Governance reforms targeting concentration risk, such as token lockups, quadratic voting, or partial delegation, may help reduce disproportionate influence without discouraging large token holders from participating. Effective design in this area can facilitate broad-based input while preserving capital efficiency \cite{ostrom}.

\noindent
More broadly, aligning governance practices with DAO-specific operational profiles (e.g. adjusting voting durations based on proposal type) may help ensure decision quality and continuity over time.
Our findings confirm that persistent low participation remains a core vulnerability. This aligns with prior studies indicating that under 10\% voter turnout can lead to oligarchic outcomes \cite{feichtinger2023hidden}. In particular, large token holders often propose and pass initiatives with minimal community input.
\noindent
We showed that DAOs with more equitable token distribution and moderate voting windows (3–14 days) achieve higher median approval rates and more balanced proposer activity. Hence, introducing tiered quorums or partial delegation (as in Liquid Democracy) could further diversify proposer authority.

\subsection{Comparisons with Existing Literature}
\label{subsec:discussion_literature}

\noindent
The observed patterns of partial decentralisation are consistent with prior research on governance concentration in DAOs \cite{feichtinger2023hidden, wang2019decentralized}. While many DAOs aim to implement community-led models, token distribution and proposer activity often remain uneven. This study builds on previous work by quantifying how treasury size, participation rates, and voting structure relate to sustainability indicators.

\noindent
The moderate association between treasury size and participation mirrors earlier findings by \cite{faqir2021comparative}, where financial capacity alone did not guarantee user engagement. These findings reinforce the view that sustainable governance depends on multiple interacting factors, not isolated metrics.

\subsection{Limitations and Opportunities for Future Research}
\label{subsec:discussion_limitations}

\noindent
Few limitations should be noted. First, the dataset includes only DAOs meeting certain on-chain activity thresholds, excluding organisations with limited or off-chain governance. Second, the analysis provides a snapshot as of April 2025; DAO structures and participation trends may evolve over time.

\noindent
Future work could incorporate longitudinal analysis to observe governance changes over time, introduce weighting schemes based on governance outcomes, or expand coverage to off-chain processes through community forums and governance platforms. Analysing the gap between the realities of DAO governance and its ideals \cite{stewardship} represents another avenue for future work. Extending the pipeline to non-EVM-compatible chains, or integrating data across bridging protocols, would improve generalisability across DAO ecosystems. Finally, refining score thresholds and incorporating adaptive metrics could make the framework more responsive to DAO-specific use cases, such as fast-moving DeFi projects or socially-driven communities.

\section{Threats to Validity}
\label{sec:threats_validity}

\subsection{Construct Validity}
\label{subsec:threats_construct}

\noindent
\textbf{Definition of KPIs.}
The four KPIs provide a structured view of DAO governance but are proxies for broader qualities. Off-chain deliberation, community sentiment, and informal leadership are not captured in on-chain data. As such, participation and voting metrics may underestimate engagement in DAOs that rely on off-chain activity.

\noindent
\textbf{Boundaries of KPI Thresholds.}
Thresholds (e.g.\ 10\% and 40\% for participation; \$100 million and \$1 billion for treasury size) are based on empirical patterns and conceptual rationale. However, small differences (e.g.\ 9\% vs.\ 10\%) can shift category placement. Token-based decentralisation thresholds may also not apply uniformly across DAO types, introducing subjectivity into classification.

\subsection{Internal Validity}
\label{subsec:threats_internal}

\noindent
\textbf{Data Completeness.}
The dataset includes DAOs with active on-chain governance, excluding those led off-chain. This reduces reliance on aggregator dashboards but may bias the sample toward more formal or transparent governance models, influencing KPI trends.

\noindent
\textbf{Measurement Accuracy.}
Node queries and event decoding can be affected by parsing errors, outdated ABIs, or contract anomalies. Despite validation (e.g.\ \texttt{ProposalCreated}/\texttt{Executed} checks), some modules deviate from expected schemas. We addressed this via schema-based normalization and curated data, reaching 99.8\% coverage.

\subsection{External Validity}
\label{subsec:threats_external}
\noindent
\textbf{Generalisability of Findings.}
Our analysis focuses on Ethereum and EVM-compatible networks, which follow ERC-20/721 governance models. Governance in Tendermint-, Cosmos-, or Substrate-based DAOs may differ, so findings may not generalise beyond EVM contexts.

\noindent
\textbf{Temporal Context.}
The dataset captures DAO governance as of April 2025. Given evolving rules, tokenomics, and participation, KPI scores may change. A longitudinal approach would better reflect governance stability and change over time.

\subsection{Reliability}
\label{subsec:threats_reliability}
\noindent
\textbf{Reproducibility of the Pipeline.}
The analysis relies on third-party services (e.g.\ Infura, Alchemy, Etherscan), which may face downtime or updates. Script or API changes can affect future runs. While version control and pinned contracts were used, replication may still be impacted by ABI or ecosystem changes.

\noindent
\textbf{Thresholding and Scoring Variations.}
An equal-weighted scoring scheme was used to reduce bias, but alternative weights or finer scoring could shift results. Open-source code and clear definitions support replication and comparison across models.

\bigskip
\noindent
While the study ensures robustness through direct data collection and defined KPIs, some limitations remain: threshold trade-offs, exclusion of off-chain activity, and limited generalisability to EVM-based DAOs. Still, the data supports strong links between governance sustainability and participation, decentralisation, financial robustness, and procedural efficiency. Future work integrating off-chain and multi-chain data can expand coverage and address these gaps.

\section{Conclusion}
\label{sec:conclusion}

\noindent
This study introduced a data-driven framework for evaluating DAO governance, combining four empirically grounded KPIs: Network Participation, Accumulated Funds, Voting Mechanism Efficiency, and Decentralisation, into a unified analytical model. Analysis of on-chain data across a diverse set of 50 DAOs revealed that higher sustainability scores tend to be associated with broader participation, decentralised control, and balanced financial and procedural structures. These findings reinforce the view that sustainability is shaped by the interaction of social, economic, and procedural factors, rather than any single attribute. The framework offers a replicable basis for comparative analysis and may support both academic research and practitioner decision-making in decentralised governance. Evaluating DAOs as complex socio-technical systems, rather than through isolated metrics, provides a more accurate account of their governance dynamics. Future work can refine this model through longitudinal evaluation, off-chain data integration, and cross-chain extensions.

\balance
\bibliographystyle{IEEEtran}
\bibliography{references}

\end{document}